# Distinctive ionic transport of freshly excised human epileptogenic brain tissue

**David Emin**,[1,a)] **Aria Fallah**,[2] **Noriko Salamon**,[3] **Gary Mathern**,[2] **and Massoud Akhtari**[4]

[1] Department of Physics and Astronomy, University of New Mexico, Albuquerque, New Mexico 87131, USA

[2] Department of Neurosurgery and Pediatrics, David Geffen School of Medicine, University of California at Los Angeles, Los Angeles, California 90095, USA

[4] Department of Radiology, David Geffen School of Medicine, University of California at Los Angeles, Los Angeles, California 90095, USA

[4] Semel Institute for Neuroscience and Human Behavior, David Geffen School of Medicine, University of California at Los Angeles, Los Angeles, California 90095, USA

a) **Author to whom correspondence should be addressed**: emin@unm.edu

## ABSTRACT

Epileptogenic lesions have higher concentrations of sodium than does normal brain tissue. Such lesions are palpably recognized by a surgeon and then excised in order to eliminate epileptic seizures with their associated abnormal electrical behavior. Here we study the frequency-dependent electrical conductivities of lesion-laden tissues excised from the brains of epilepsy patients. The low-frequency (< 1000 Hz) conductivity of biological tissue primarily probes extracellular solvated sodium-cations traveling parallel to membranes within regions bounded by blockages. This conductivity rises monotonically toward saturation as the frequency surpasses the rate with which diffusing solvated sodium cations encounter blockages. We find that saturation occurs at dramatically higher frequencies in excised brain tissue containing epileptogenic lesions than it does in normal brain tissue. By contrast, such an effect is not reported for tumors embedded in other excised biological tissue. All told, epileptogenic lesions generate frequency-dependent conductivities that differ qualitatively from those of both normal brain tissues and tumors.

Ionic motion is central to biological processes. For almost a century this knowledge motivated studies of the room-temperature charge transport of human and animal biological matter over a very wide range of frequencies, 1 Hz to $10^{11}$ Hz.[1,2] In detail, at the highest frequencies (in the so-called $\gamma$-regime, $10^8$ to $10^{11}$ Hz) frequency-dependent charge transport is dominated by reorienting water molecules which pervade biological tissue. At intermediate frequencies (in the so-called $\beta$-regime, $10^3$ to $10^8$ Hz) frequency-dependent conductivities are dominated by polarizations of membranes' bilayers and their interfaces with surrounding ions. Less attention has been focused on the low-frequency domain (the so-called $\alpha$-regime, < $10^3$ Hz). In the $\alpha$-regime the frequency-dependent conductivity is dominated by extracellular solvated ions moving nearly parallel to membranes. Such low-frequency conductivities are altered by the inclusion of diseased matter (e.g. tumors) in biological tissues.

The $\alpha$-regime charge transport of normal brain tissue is predominantly that of extracellular sodium cations solvated in the brain's water.[3,4] In particular, extracellular sodium concentrations are very much greater than intracellular concentrations; e.g. in normal rat brain, the concentration of extracellular sodium is 14 times that of intracellular sodium.[5] Furthermore, extracellular diffusion of solvated sodium cations occurs with reorientation of the surrounding water molecules, which can be monitored by hydrogens' diffusion-MRI.[6]

The frequency dependence of the conductivity is generally determined by blockages which limit the spatial domain within which carriers can move.[7] In particular, the conductivity falls from its intrinsic value when carriers encounter blockages. Thus, the ionic conductivity of normal tissue (e.g. brain tissue of sacrificed healthy rats) generally increases monotonically with increasing frequency to a saturation value as diffusing ions increasingly avoid cellular blockages.[3,5-8] This saturation value, typically reached between 50 and 100 Hz in healthy tissue, therefore measures the *intrinsic* ionic conductivity of normal tissue.[3,5-8] The corresponding separations between $\alpha$–regime blockages in normal brain tissue are of the order of $10^{-5}$ m. These lengths are between those probed by diffusion-MRI and pathology studies.

Here we investigate the conductivities between 6 Hz and 1000 Hz of approximately 1 $cm^3$ fresh samples taken from tissues resected during brain surgeries to excise lesions from 21 recent human epilepsy patients. Excised material from each patient is measured at room temperature. Redacted material from each patient 1) undergoes pathology examination, 2) diffusion-MRI (i.e. spin echo nuclear magnetic resonance, NMR) measurement of its water molecules' hydrogen nuclei, and 3) measurement of its $\alpha$-regime frequency-dependent conductivity. We also note that *in vivo* MRI measurements on $^{23}Na$ find that the sodium concentrations of epileptogenic lesions are greater than those of normal brain tissue.[9]

Our resected tissue is generally a mixture of an epileptogenic lesion and some surrounding and intermingled normal tissue. Samples taken from resected tissue for use in conductivity measurements are therefore also mixtures of an epileptogenic lesion and some ancillary normal brain tissue.

In a small percentage of early samples (< 5%) transport is through normal tissue modified by the release of a significant fraction of sodium cations from traps within lesions.[7] A distinctive signature of this unusual situation is that the frequency-dependent conductivity $\sigma(\omega)$ *falls* as it approaches its saturation value, i.e., its partial derivative with respect to frequency becomes negative, $\partial\sigma(\omega)/\partial\omega < 0$.[7,8]

Recent excised material is estimated to contain a somewhat smaller fraction of normal tissue than did prior excised tissue. Indeed, low-frequency charge transport in all but one sample taken from recently excised tissues appears to be through its lesions. Distinctively, the $\alpha$-region ionic conductivities of these samples saturate at much higher frequencies than does that for normal brain tissue. In particular, we find conductivities that rise significantly with increasing frequency as they approach 1000 Hz. Thus, such behavior, rather than saturation below 50-100 Hz, indicates ionic transport within epileptogenic lesions.

This paper is organized simply. We first derive a formula for the $\alpha$-regime frequency-dependent conductivity of mixtures comprising an epileptogenic lesion along with normal brain tissue. We then present representative examples of the frequency-dependent conductivities of samples of brain tissue excised from our epilepsy patients. For comparison, we also display examples of the frequency-dependent conductivities of normal grey-matter and white-matter brain tissue. We finally compare our results for epileptogenic lesions with those of common tumors. In particular, tumors and epileptogenic lesions are both associated with enhanced sodium concentrations.[1,9-11] However, tumors primarily increase the magnitudes of their tissues' $\alpha$-regime conductivities. By contrast, epileptogenic lesions dramatically increase the frequencies at which their conductivities saturate. This finding indicates that the membranes in lesions that guide $\alpha$-regime transport of extracellular sodium-cations have many more blockages than are in normal brain tissue and common tumors. Simply put, the integrity of the membranes that control extracellular sodium-cation conductivity in epileptogenic lesions appears diminished relative both to those of normal brain tissue and tumors.

Excised tissues' intrinsic ionic conductivities are essentially those of ions solvated in water. That is, the $H_2O$ molecules surrounding an ion are oriented in accord with its charge.[3,4] In particular, a neighboring water molecule's oxygen atom tends to be closer to a solvated sodium-cation than are its hydrogen atoms.

The motion of a solvated ion is then coordinated with motions of the surrounding water molecules. Distinctively, the absence of the severe constraints on the positions and orientations of water's $H_2O$ molecules enables them to collectively pass through a multitude of configurations without significant energetic variations.[12] As a result, the diffusion constants of ions solvated in water as well as its self-diffusion constant are all of the same order, $D \approx 10^{-9}$ m$^2$/s.[13]

Our experiments confirm this estimate. In particular, proton-diffusion-MRI measurements of excised human brain tissue yield $D$ values of the order of $10^{-9}$ m$^2$/s.[3,6] Furthermore, the diffusion constants of sodium cations deduced from conductivity measurements on excised human brain tissue are also of the order of $10^{-9}$ m$^2$/s.[3,6]

As before, we model samples used in our conductivity measurement most simply as just being composed of a lesion and subsidiary normal brain tissue.[7,8] The fraction of a sample comprising a lesion is denoted by $f$ with $1 - f$ denoting the fraction of the sample composed of normal tissue. Then, with ionic motion restricted either to the lesion or to normal tissue, the frequency-dependent conductivity is:

$$\sigma(\omega) = f\sigma_l(\omega) + (1-f)\sigma_n(\omega). \quad (1)$$

The frequency-dependent conductivity of each structural component is a sum of contributions from different mechanisms which we designate by the index $i$.[1,2] Each contribution is the product of a coefficient $a_i$ and a frequency-dependent factor having the form $[\omega^2/(R_i^2 + \omega^2)]$, where $R_i$ denotes the rate with which ions involved in the $i$-th mechanism encounter blockages which restrict their motion [c.f. Eq. (15) of Ref. 8].[8] With increasing frequency this frequency-dependent factor rises monotonically from a very small value, $(\omega/R_i)^2 << 1$ for $\omega << R_i$, toward 1 for $\omega >> R_i$. Thus, with increasing frequency the conductivity rises in steps associated with successively larger values of $R_i$:

$$\sigma_l(\omega) = \sum_i a_{l,i}\left(\frac{\omega^2}{R_{l,i}^2 + \omega^2}\right), \quad (2)$$

and

$$\sigma_n(\omega) = \sum_i a_{n,i}\left(\frac{\omega^2}{R_{n,i}^2 + \omega^2}\right). \quad (3)$$

The conductivity of biological tissue is divided into three broad frequency domains.[1,2] The lowest frequency domain (below roughly 1000 Hz) is termed the $\alpha$-domain. In this regime, the frequency-dependent conductivity results from extracellular solvated ions moving parallel to membranes within regions limited by blockages. In normal brain tissue the separation between such blockages is of the order of $10^{-5}$ m. Disease- and injury-induced tissue disorganization may shorten these separations thereby increasing the rates with which solvated ions encounter blockages, $R_i$. In this manner, the low-frequency ($\alpha$- domain) conductivity probes the integrity of a tissue's structure.

At higher frequencies (roughly $10^3$ Hz to $10^8$ Hz), in the $\beta$-domain, the conductivities of biological tissue are dominated by the polarizations of semi-porous membranes and their interfaces with adjacent ions.[1,2] The corresponding blockages have separations that are comparable to membrane widths, about $10^{-8}$ m. At the highest frequencies ($10^8$ Hz to $10^{10}$ Hz), in the $\gamma$-domain, conductivities are dominated by reorientation of water molecules that pervade most biological tissue.[1,2] The length scale associated with reorienting a water molecule is simply its size, about $10^{-10}$ m.

Our studies focus on the low-frequency conductivity, < 1000 Hz. Therefore, we now retain only the contributions to the lesion-laden and normal-tissue conductivity, Eqs. (2) and (3), which predominate in the low-frequency limit. These contributions correspond to the smallest values of $R_{l,i}$ and $R_{n,i}$, respectively. Inserting these contributions into Eq. (1) yields:

$$\sigma(\omega) = f\sigma_l\left(\frac{\omega^2}{R_l^2+\omega^2}\right) + (1-f)\sigma_n\left(\frac{\omega^2}{R_n^2+\omega^2}\right), \quad (4)$$

where $R_l$ and $R_n$ respectively denote the smallest values of $R_{l,i}$ and $R_{n,i}$. In obtaining Eq. (4), we also noted that the conductivities of a lesion and normal tissue revert to their intrinsic dc values, $\sigma_l$ and $\sigma_n$, in the limit that their carriers encounter no blockages, $R_l = 0$ and $R_n = 0$, respectively. The characteristic separations between blockages in lesions and normal brain tissue are $2(2D/R_l)^{1/2}$ and $2(2D/R_n)^{1/2}$.

With tissues' carrier concentrations being sufficiently low, each of the intrinsic dc conductivities in Eq. (4) is simply proportional to the corresponding tissues' carrier density. Tissues' carriers are primarily sodium cations. Furthermore, extracellular sodium concentrations are very much greater than intracellular concentrations; e.g. in normal rat brain, the concentration of extracellular sodium is 14 times that of intracellular sodium.[5] In addition, MRI measurements with $^{23}$Na indicate significantly higher concentrations in epileptogenic lesions than in normal tissue.[9] Therefore, epileptogenic lesions presumably have exceptionally high densities of extracellular solvated sodium cations.

These effects are illustrated in Fig. 1. Figure 1 plots the relative conductivity, $\sigma(\omega)/\sigma_n$, from Eq. (4) versus the logarithm of the applied frequency $\omega$ for curves with different fractions of epileptogenic lesions, $f$. These plots presume that the exceptionally high sodium concentration and the disorganization of epileptogenic lesions respectively increase their intrinsic conductivity and the rate with which their carriers encounter blockages relative to those in normal tissue: $\sigma_l > \sigma_n$ and $R_l >> R_n$. Distinctively, the conductivity's frequency dependence shifts dramatically to higher frequencies as $f$ is increased from zero. Thus, this feature of the frequency-dependent conductivity indicates the presence of epileptogenic lesions.

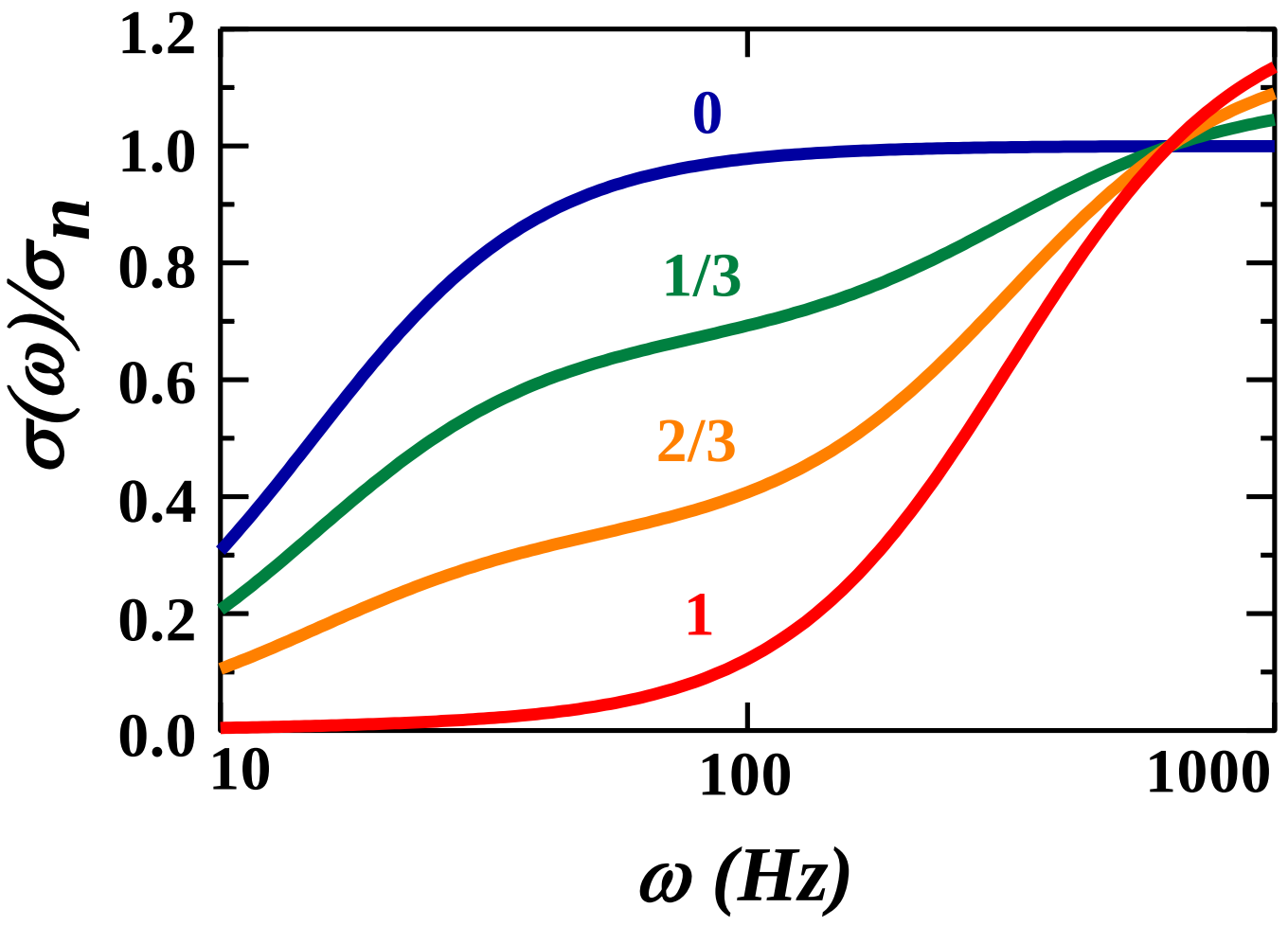

Fig. 1. The frequency-dependent conductivity $\sigma(\omega)$ of Eq. (4) is plotted in units of $\sigma_n$ versus the applied frequency $\omega$ for $f$= 0, 1/3, 2/3 and 1, the fraction of a sample that is composed of epileptogenic lesions. The rates at which sodium cations are blocked within normal brain tissue and an epileptogenic lesion are herein taken to be $R_n$ = 15 Hz and $R_l$ =1000 Hz with $\sigma_l$ = 1.25 $\sigma_n$. Thus, the blue ($f$ = 0) and red ($f$ = 1) curves respectively denote the relative conductivities of normal brain tissue and epileptogenic lesions.

We now compare these findings with our measurements of the frequency-dependent conductivities of 1 $cm^3$ samples taken from larger (from a few $cm^3$ up to over 300 $cm^3$) lesion-containing tissues excised during brain surgeries performed over the last five years on 21 epilepsy patients. This number significantly increases the database of studied patients from its prior value, 67, which was accumulated since these measurements began in 2007. Eighteen of these recent patients were infants or children while three of these patients were over 40 years of age. Twelve of these patients were male and nine of these patients were female. The lesions were resected from different parts of the brains of patients with a variety of epilepsy etiologies. Seven patients had been diagnosed with cortical dysplasia; Three patients were diagnosed with perinatal ischemia; Two patients each received a diagnosis of either Rasmussen, Chaslin's gliosis, tuberous sclerosis, or sclerosis; Three patients received a diagnosis of either gliosis, chronic trauma, or cavernous hemangioma.

Figures 2 through 5 show the frequency-dependent conductivities of four representative examples of our findings. For comparison with normal mammalian brains, the frequency-dependent conductivities of normal bovine grey matter and normal bovine white matter are depicted in Figs. 6 and 7, respectively.

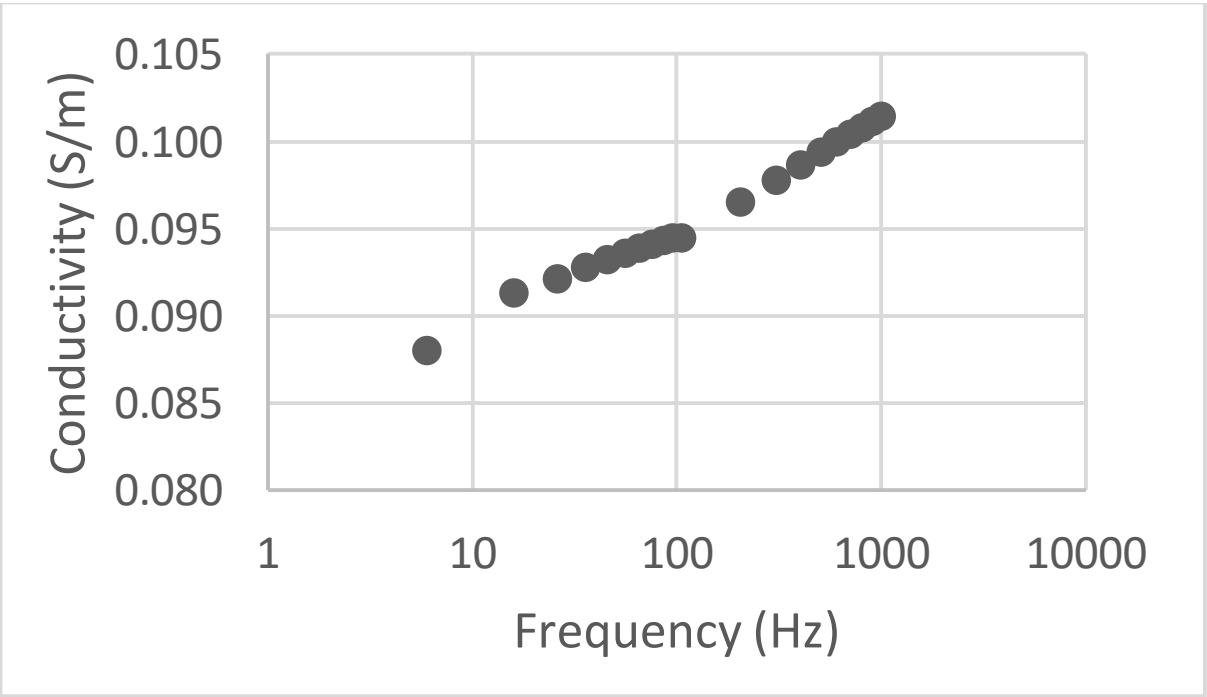

Fig. 2. The conductivity versus frequency of a sample extracted from tissue excised from the right peri-insular region of an eleven month-old female diagnosed with cortical dysplasia.

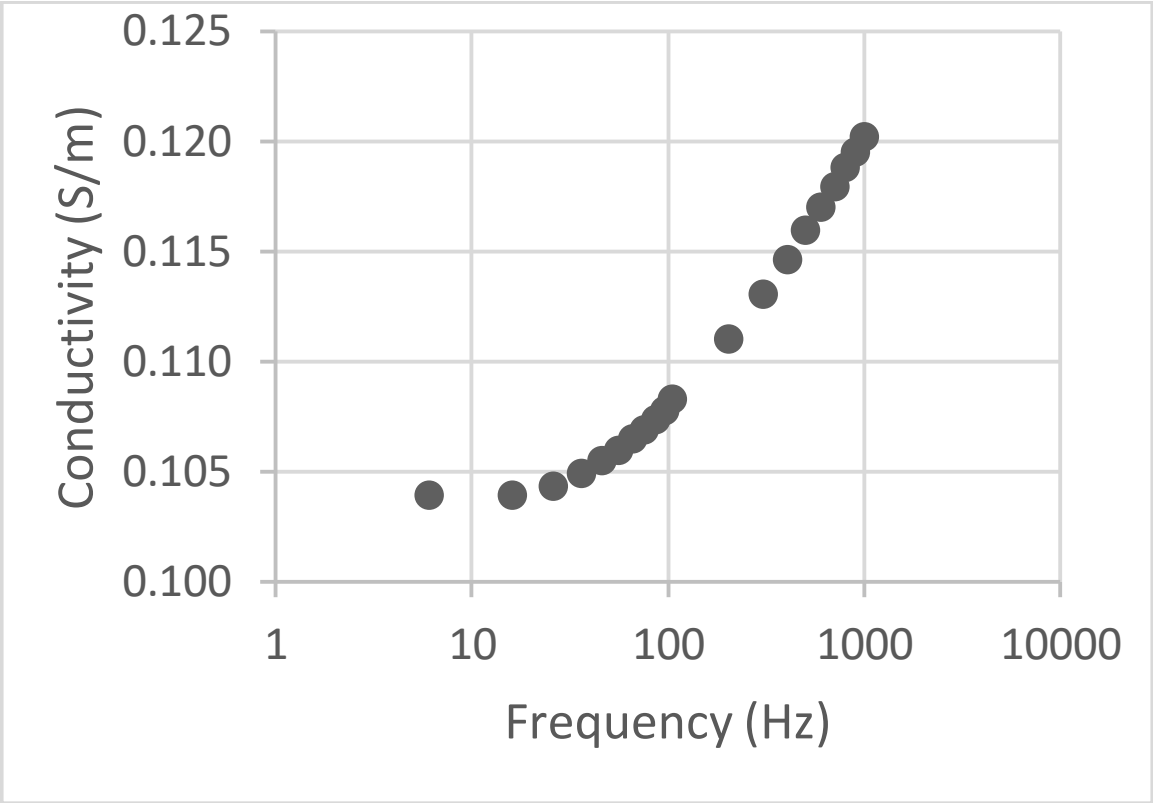

Fig. 3. The conductivity versus frequency of a sample extracted from tissue excised from the left occipital lobe of a seven-year old male diagnosed with cortical dysplasia.

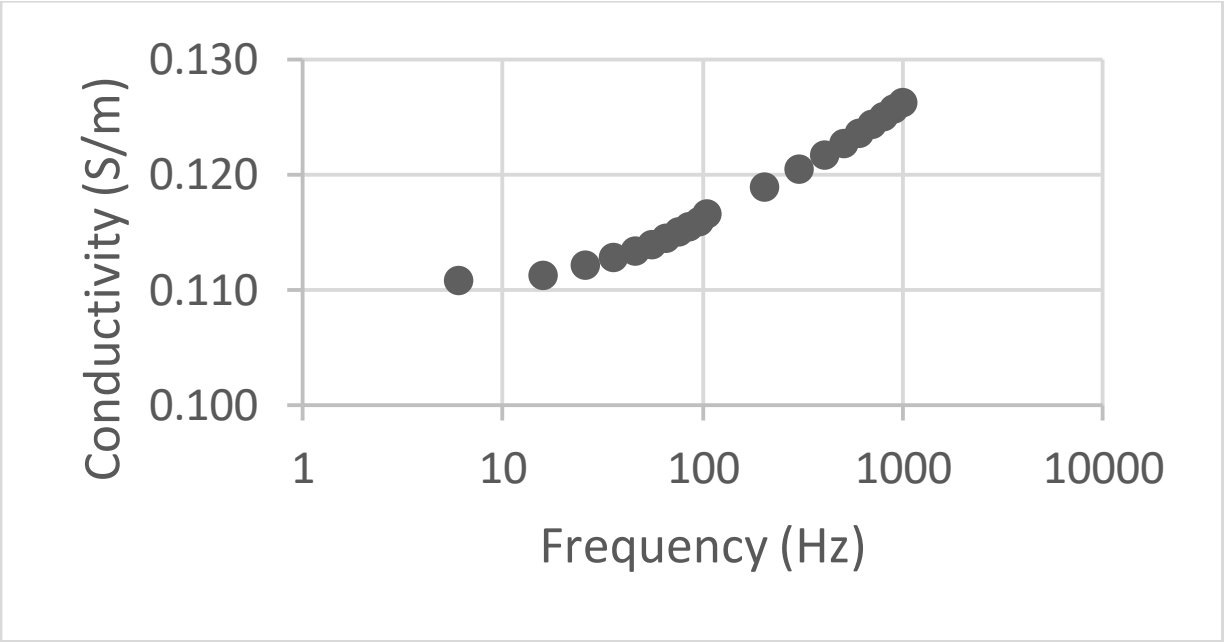

Fig.4. The conductivity versus frequency of a sample extracted from tissue excised from the left insula of a three year-old male diagnosed with (infantile spasm) cortical dysplasia.

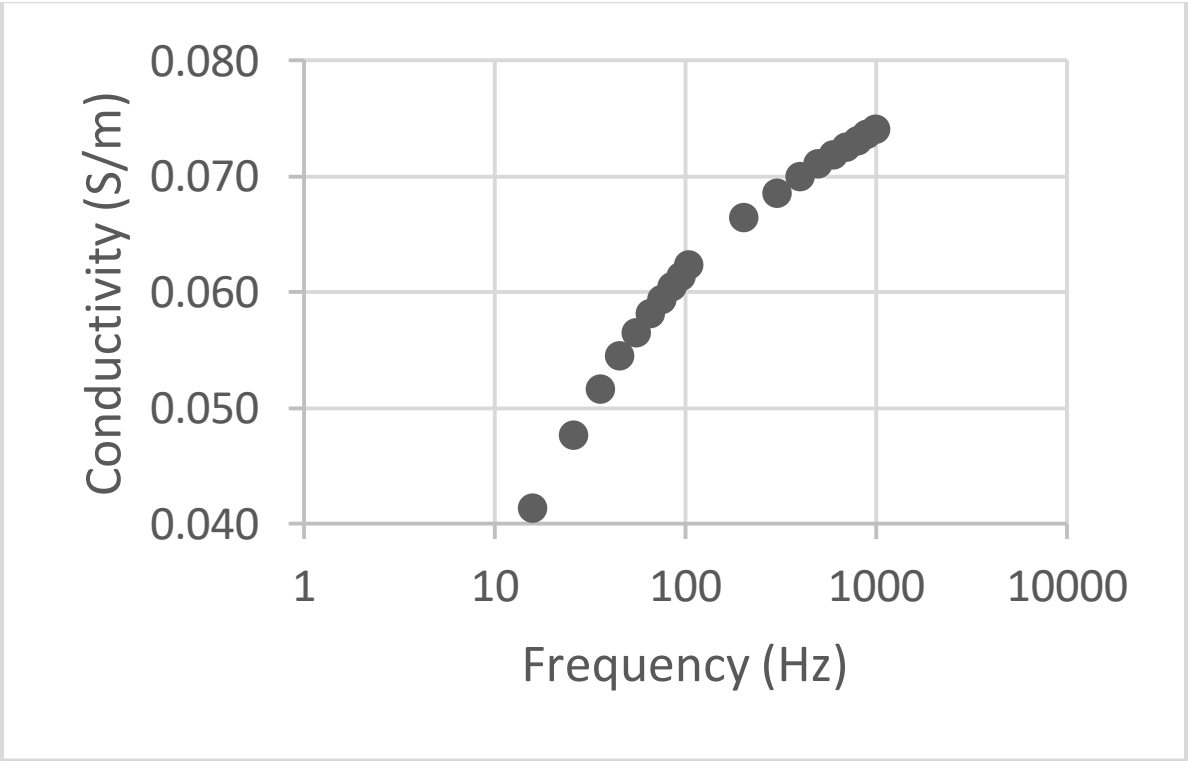

Fig. 5. The conductivity versus frequency of a sample extracted from tissue excised from the left posterior temporal lobe of a twenty-two month-old female who suffered from a stroke.

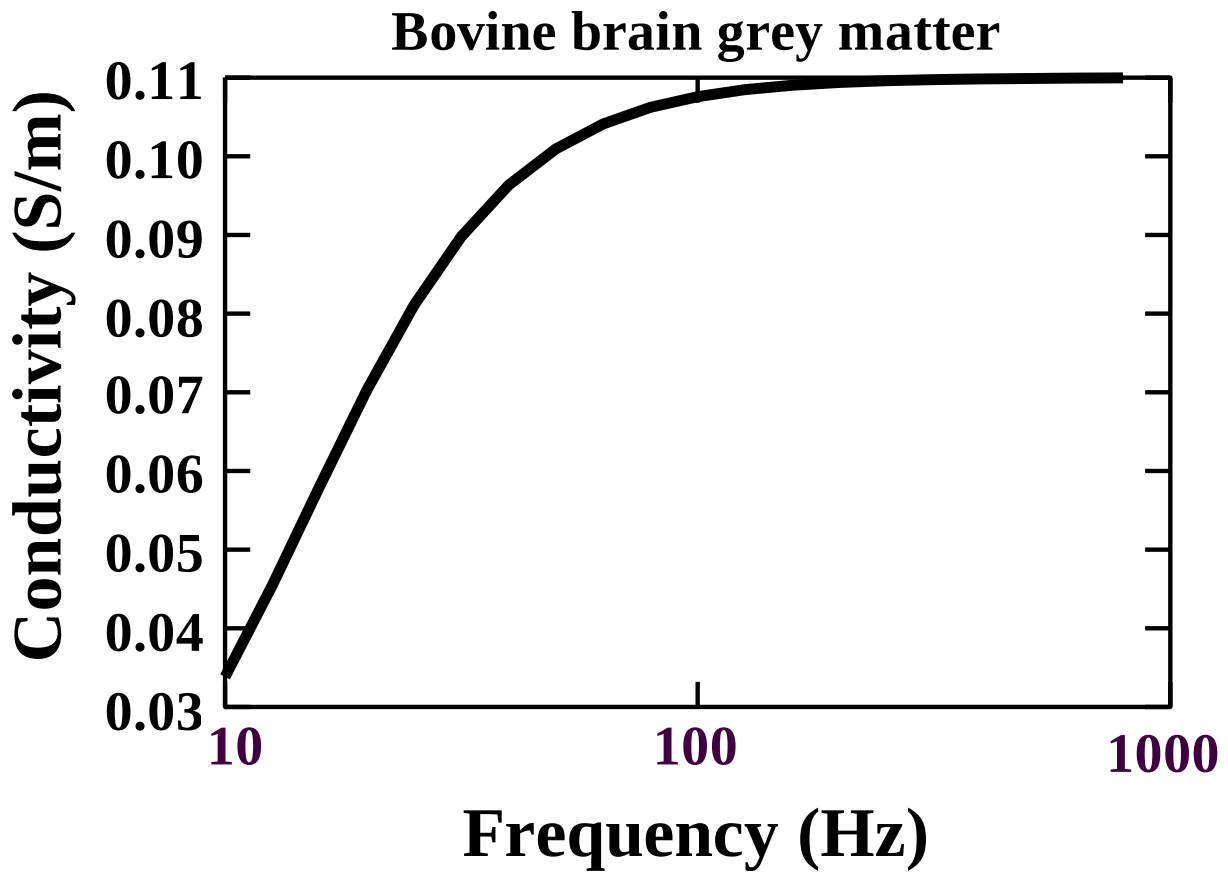

Fig. 6. The conductivity versus frequency is shown for grey matter of a healthy bovine brain. The curve is our fit of our Eq. (4) to data from Ref. 14 with $f = 1$, $\sigma_n = 0.11$ S/m, and $R_n = 15$ Hz.

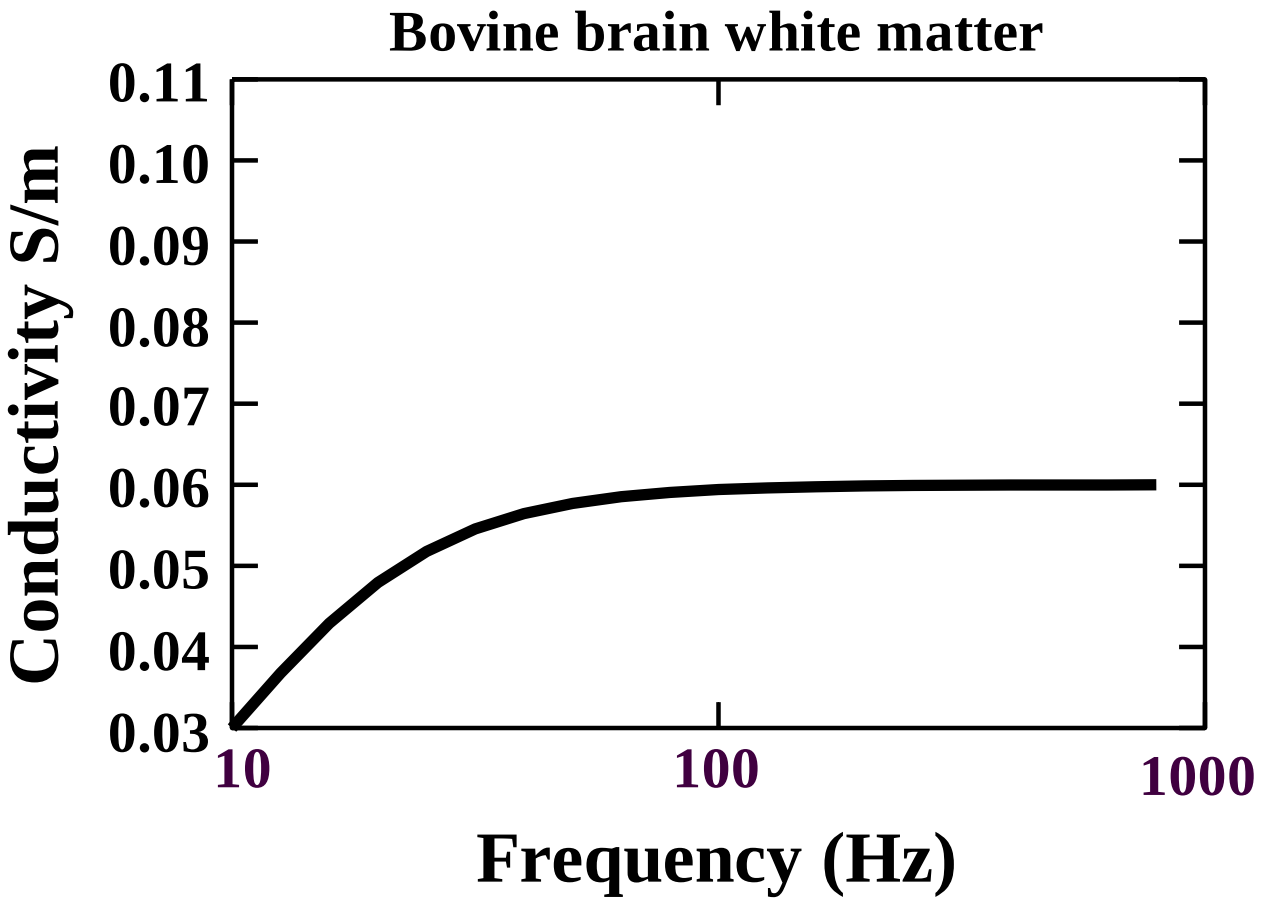

Fig. 7. The conductivity versus frequency is shown for white matter of a healthy bovine brain. The curve is our fit of our Eq. (4) to data from Ref. 14 with $f = 1$, $\sigma_n = 0.06$ S/m, and $R_n = 10$ Hz.

The salient feature of the conductivities of excised lesion-containing samples from epilepsy patients, shown in Figs. 2-5, is that their rise with increasing frequencies extends well above 100 Hz. In dramatic contrast, the conductivities of normal brain tissue, illustrated in Figs. 6 and 7, tend toward saturation below 100 Hz. All told, the frequency dependencies of the conductivities of samples containing epileptogenic lesions are qualitatively different from the frequency dependencies of normal brain tissue.

In addition, the conductivities of our samples at 1000 Hz are at least comparable to the saturation values of normal mammalian brain tissues. This result suggests higher $\alpha$-regime saturation conductivities for our lesion-containing samples than for normal brain tissue. Thus, our lesion-containing samples contain higher densities of carriers, presumably sodium cations, than exist in normal brain tissue. This conclusion is consistent with $^{23}$Na-MRI measurements finding that epileptogenic lesions have exceptionally high concentrations of sodium atoms.[9]

In summary this paper addresses the room-temperature $\alpha$-regime frequency-dependent conductivities of samples taken from resected tissue containing lesions excised from the brains of 21 epilepsy patients. At the low frequencies that define the $\alpha$-regime, 1 Hz to 1000 Hz, charge transport is dominated by extracellular ions that move nearly tangentially along membranes.

Extracellular ions are generally solvated in the water that pervades brain tissue. As such, their intrinsic room-temperature diffusion constants are about $10^{-9}$ $m^2$/sec.[13] Our diffusion MRI and conductivity measurements are consistent with this estimate of solvated ions' diffusion constant.

In addition to solvated ions' diffusion constant, the conductivity is proportional to 1) the density of diffusing ions and 2) a frequency-dependent factor which depends upon the rate at which these ions encounter blockages. In other words, this frequency-dependent factor measures the integrity of the membranes that guide the flow of extracellular solvated ions. Both of these two factors can be sensitive to the presence of diseased tissue.

Epileptogenic lesions and tumors both have abnormally high concentrations of sodium atoms.[1,9-11] High concentrations of sodium atoms imply high densities of their diffusing solvated sodium cations, presumably the primary extracellular ionic charge carriers. Thus, epileptogenic lesions and tumors are both associated with carrier densities that exceed those of normal tissue. By itself, this effect simply raises the magnitudes of the $\alpha$–regime conductivities of lesion-laden and tumor-laden tissues relative to those of normal tissue.

Simple increases of the $\alpha$-regime conductivities are reported for tumor-laden tissue.[15-19] For example, Fig. 3 of Ref. 19 shows that the presence of tumors from muscles of over 15 mice only significantly alters their conductivity in the $\alpha$-regime, $< 1000$ Hz. Furthermore, these conductivities are only larger in magnitude than those of normal mouse-muscle tissue. Tumors do not change the frequency dependences of the measured $\alpha$-regime conductivities. To rationalize these results, we note that solid tumors' primary constituents are stroma which mimic surrounding normal tissue albeit with somewhat enhanced water and ion concentrations.[20]

The overriding effect of epileptogenic lesions is that their inclusion in excised brain tissue dramatically increases the frequencies at which their conductivities saturate as functions of frequency. This feature is obvious upon comparing Figs. 2-5 with Figs. 6 and 7. As shown in Fig. 1, this effect results from epileptogenic lesions introducing blockages to the extracellular flow of solvated sodium-cations. Thus, the integrity of the membranes which guide $\alpha$–regime ion flow in epileptogenic lesions is much less than that in normal brain tissue. Distinctively, the lengths probed by the $\alpha$–regime conductivity are between those probed by diffusion-MRI measurements and pathology examinations.

All told, we find that the inclusion of epileptogenic lesions in excised brain tissues introduces striking changes to their $\alpha$-regime frequency-dependent conductivities. The resulting frequency-dependent conductivities are qualitatively distinct from those of tumors as well as from those of normal brain tissues.

This study was supported by NIH R21 NS108247-01 and the Weil Fund at UCLA Semel Institutes for Neuroscience and Human Behavior.

The data described in this article are openly available within the cited articles.

## REFERENCES

1. R. Pethig and D. B. Kell, Phys. Med. Biol, **32**, 933-970 (1987).

2. H. P. Schwan, Adv. Biol. Phys. **5**, 147-209 (1957).

3. D. Emin, M. Akhtari, B. M. Ellingson, and G. W. Mathern, AIP Advances **5**, 087143 (2015).

4. B. Hirbar, T. Southhall, V. Vlacky, and K. A. Dill, J. Am. Chem. Soc. **124**, 12302 (2002).

5. J. A. Goodman, C. D. Kroenke, G. L. Bretthorst, J. H. Ackerman, and J. J. Neil, Magnetic Resonance in Medicine **53**, 1040-1045 (2005).

6. M. Akhtari, D. Emin, B. M. Ellingson, D. Woodworth, A. Frew, and G. W. Mathern, J. Appl. Phys. **119**, 064701 (2016).

7. D. Emin, M. Akhtari, A. Fallah, H. V. Vinters, and G. W. Mattern, J. Appl. Phys. **122**, 154701 (2017).

8. D. Emin, N. Salamon, W. Yong, A. Frew, G. Frew, G. Mathern, and M. Akhtari, AIP Advances **11**, 045118 (2021).

9. B. Ridley, A Marchi, J. Wirsich, E. Soulier, S. Confort-Gouny, L. Schad, F. Bartolomei, J-P Ranjeva, M. Guye, and W. Zaaraoui, NeuroImage **157**, 173-183 (2017).

10. R. Damadian and F. W. Cope, Physiol. Chem. Phys, **6**, 309-322 (1974).

11. T. B. Pool, I. L. Cameron, N. K. R, Smith, and R. L. Sparks, *The Transformed Cell*, I. L. Cameron and T. B. Pool, eds. (Academic, New York, 1981).

12. H. Fröhlich, *Theory of Dielectrics*, (Oxford, London, 1958) Chap. IV Sec. 17.

13. J. Frenkel, *Kinetic Theory of Liquids* (Dover, New York, 1955), Chap. 4 Sec. 4.

14. S. Gabriel, R. W. Lau, and C. Gabriel, Phys. Med. Biol. **41**, 2271-2293 (1996).

15. H. Fricke and S. Morse, J. Cancer Res. **10**, 340-346 (1926).

16. P. A. Bottomley and E. A. Andrews, Phys. Med. Biol. **23**, 630-643 (1978).

17. B. Singh, C. W. Smith, and R. Hughes, Med. Biol. Eng. Comput. **17**, 45-60 (1979).

18. K. R. Foster and J. L. Schepps, Microwave Power **16**, 108-119 (1981).

19. J. A. Rogers, R. J. Sheppard, E. H. Grant, N. M. Bleehen, and D. J. Honess, Br. J. Radiol. **56**, 335-338 (1983).

20. J. L. Connolly, S. J. Schnitt, H. H. Wang, J. A. Longtime, A. Dvorak, and H. F. Dvorak, *Holland-Frei Cancer Medicine* 6$^{th}$ Ed., D. W. Kufe, R. E. Pollock, R. R. Weichselbaum, C. Bast, Jr, T. S. Gansler, J. F. Holland, and E. Frei, III, editors (BC Decker, Hamilton, 2003) Chap. 35 Sec. 1.